# Propagation of videopulse through a thin layer of two-level atoms possessing permanent dipole moments

Sergei O. Elyutin[1] and Andrei I. Maimistov

*Department of Physics, Department of Solid State Physics, Moscow Engineering Physics Institute, Moscow 115409*

**ABSTRACT**

The excitation of a thin layer of two-level permanent dipole moment atoms by ultimately short (less than field oscillation period) electromagnetic pulses (videopulse) is observed. The numerical analysis of matter equations free of rotating wave approximation and relaxation reveals a strong affect of local field and Stark effect on temporal behavior of transmitted field. Specifically it is demonstrated that a dense film irradiated by videopulse emits a short response with a delay much longer even than the characteristic cooperative time of atom ensemble. It is supposed that the local field in the thin layer of permanent dipole atoms is able to re-pump the atomic subsystem. The close analogy to nonlinear pendulum motion is discuused.

## 1. INTRODUCTION

Because of an ultimately short duration of videopulse the electric field polarization cannot change polarity in the time of pulse action[1-3]. Such pulses represent bursts of energy of the field, for which the concepts of carrier frequency and wavelength are no longer applicable[4-6]. The conventional for nonlinear optics of quasimonochromatic pulses slow envelope approximation and rotating wave approximations[7,8] should be rejected or at least specified to account the next orders of perturbation theory. The propagation of ultimately short pulses including videopulse in a two-level medium was investigated in a typical for many resonance media situation, when the diagonal elements of the dipole operator are zero[9,10]. However, there are media (molecular gases, adsorbed atoms, quantum dots arrays) where the linear Stark effect takes place[11-13]. In these cases, the diagonal elements of dipole transition operator between resonance levels are non-zero.

Analogous to Kerr media, the media with non-zero diagonal elements of dipole moment can be called Stark media[14-15]. The interaction of a few cycle electromagnetic pulse with Stark medium with no account of propagation effects was considered in[16]. At the same time, the lift of slow amplitude approximation supplemented with unidirectional wave approximation makes the problem completely integrable and permits to find steady-state solutions in soliton form[17-19]. Further analysis revealed a new solitary electromagnetic object, the non-zero breather[20,21].

In this paper a thin layer of permanent dipole atoms with a thickness of the order of a wavelength placed on an interface of two dielectric media represents a simple example of 2D system, which optical properties has been intensively studied[22-26]. Thin film physics is attributed by an account of microscopic field acting on resonance atoms due to dipole-dipole interaction[27-29]. The most of studies of the short pulse refraction by such nonlinear interface were carried out under the assumption

---

[1] E-mail: elyutin@mail.ru, soelyutin@email.mephi.ru

that the atoms of a film were modeled by a two-level system in one-photon resonance with the incident radiation[30,31]. The approaches worked out for conventional models were extended to another low dimension systems, such as the layer of quantum dots[32], film of three level atoms[33,34], and different type of resonance, for instance, the two-photon resonance[35,36]

The aim of this work is to demonstrate by numerical modeling, how an ultimate shortness (about a field halfperiod) of electromagnetic "buble"[1] (videopulse) and Stark effect due to the permanent atom dipole moments become apparent in interaction of pulsed field with a specific low dimensional medium – thin layer, where local field contribution is evidently perceptible. It turned out that in such model in the absence of relaxation the trigger switching of film state can be realised under the action of inner local field. It is shown that a thin layer of dipolar atoms being irradiated with a videopulse is able to emit a short signal with a very long delay, longer than all characteristic times of the problem.

## 2. THE COUPLING EQUATION

Let a thin polarizable film of thickness $l$ lie on the interface of two dielectric media. The thickness of the atom layer is supposed to be less than a spatial extent of electromagnetic pulse. Let Z-axis be normal to the interface. For a wave polarized in the plane of interface the continuity conditions look like:

$$E_y(z=0^-,t) = E_y(z=0^+,t), \qquad (1.1)$$

$$H_x(z=0^-,t) - H_x(z=0^+,t) = \frac{4\pi}{c}\frac{\partial}{\partial t}P_{S,y}(t), \qquad (1.2)$$

where $P_{S,y} = P_y(z=0,t)$ – is a surface polarization of the film.

Let the short pulse of the TE-wave be incident on the interface from a $Z<0$ side. For simplicity, the dielectric media posses no dispersion and the electromagnetic pulse has an arbitrary duration, thus not being necessarily a quasiharmonic wave. Then the solution of Maxwell equations writes in the following form:

$$E_y(z,t) = \begin{cases} E_{in}(t-z/V_-) + E_{ref}(t+z/V_+), & z<0 \\ E_{tr}(t-z/V_+), & z>0 \end{cases} \qquad (2)$$

Here $V_\pm$ is the solitary wave (pulse) group velocity in a matter with the corresponded signs of $z$ coordinate. Due to the absence of dispersion, the group velocities are the determined permanent values characterizing the media. The continuity condition (1.1) provides the relationship

$$E_{in}(t) + E_{ref}(t) = E_{tr}(t). \qquad (3)$$

It results from Maxwell equations that for the planar symmetry

$$\frac{\partial E_y}{\partial z} = \frac{1}{c}\frac{\partial H_x}{\partial t}.$$

By using this expression, one can get from (1.2) that

$$\frac{1}{c}\frac{\partial}{\partial t}H_x(z=0^+,t) - \frac{1}{c}\frac{\partial}{\partial t}H_x(z=0^-,t) = \frac{4\pi}{c^2}\frac{\partial^2}{\partial t^2}P_{S,y}(t),$$

or

$$\frac{\partial}{\partial z}E_y(z=0^+,t) - \frac{\partial}{\partial z}E_y(z=0^-,t) = \frac{4\pi}{c^2}\frac{\partial^2}{\partial t^2}P_{S,y}(t). \tag{4}$$

Making use of expression (2) and bearing in mind that introduction of retarded time $t' = t - z/V_\pm$ and $z = z'$ specifies the derivations as following

$$\frac{\partial}{\partial t} = \frac{\partial}{\partial t'} \text{ and } \frac{\partial}{\partial z} = \frac{\partial}{\partial z'} \pm \frac{1}{V_\pm}\frac{\partial}{\partial t'},$$

equation (4) can be re-written as

$$\frac{\partial}{\partial t'}\left(E_{ref} - E_{in} + \frac{V_-}{V_+}E_{tr}\right) = -\frac{4\pi V_-}{c^2}\frac{\partial^2}{\partial t'^2}P_{S,y}(t). \tag{5}$$

Equation (5) can be integrated over time, thus providing the second necessary expression

$$E_{ref} - E_{in} + \frac{V_-}{V_+}E_{tr} = -\frac{4\pi V_-}{c^2}\frac{\partial}{\partial t'}P_{S,y}(t') = -\frac{4\pi V_-}{c^2}n_S\frac{\partial}{\partial t'}p(t'), \tag{6}$$

where $n_S$ is a surface density of dipole atoms, concerned with a bulk density $n_A$ by an approximate relation $n_A \approx n_S/l$, $p(t)$ is a polarization per one atom.

The atoms of polarizing film are affected by the field $E_{tr}(t)$, if the Lorentz field correction is not accounted, $E_{in}(t)$ is the electric field of a given incident electromagnetic wave. Excluding the reflected wave $E_{ref}(t)$ from (3) and (6), one can write the relationships, which generalized the Fresnel formula to the generally non-harmonic waves:

$$E_{tr}(t') = \frac{2V_+}{V_- + V_+}E_{in}(t') - \frac{4\pi V_- V_+ n_S}{c^2(V_- + V_+)}\frac{\partial}{\partial t'}p(t') \tag{7}$$

$$E_{tr}(t') = \frac{V_+ - V_-}{V_- + V_+}E_{in}(t') - \frac{4\pi V_- V_+ n_S}{c^2(V_- + V_+)}\frac{\partial}{\partial t'}p(t')$$

A specific model for atoms or molecules (enharmonic oscillators, resonance atoms, quantum dots etc.), constituting thin film provides necessary additional equations to obtain $p(t')$.

### 3. TWO-LEVEL ATOMS POSSESING PERMANENT DIPOLE MOMENT

Let us consider a thin film of two-level atoms, similar to one observed in a number of publications[22-26], regarding, that unlike the convenient model[7], in our case the resonance transition with the frequency $\omega_a$ is featured by both the non-diagonal and diagonal matrix elements of the dipole tran-sition operator. Polarization of medium is defined as $P = n_A p$, but polarization per atom $p$ is expressed by means of matrix of density and dipole moment operator:

$$p = \mathrm{tr}\hat{\rho}\hat{d} = \rho_{11}d_{11} + \rho_{22}d_{22} + \rho_{12}d_{21} + \rho_{21}d_{12}.$$

Taking account of normalization condition $\mathrm{tr}\hat{\rho} = \rho_{11} + \rho_{22} = 1$, one can write

$$p = \frac{1}{2}(d_{11} + d_{22}) + \frac{1}{2}(d_{11} - d_{22})(\rho_{11} - \rho_{22}) + \rho_{12}d_{21} + \rho_{21}d_{12}. \tag{8}$$

The elements of density matrix $\hat{\rho}$ satisfy the equations followed from Neiman equation

$$i\hbar \partial \hat{\rho}/\partial t' = \hat{H}\hat{\rho} - \hat{\rho}\hat{H}$$

for the chosen Hamiltonian

$$\hat{H} = \frac{\hbar\omega_a}{2}\begin{pmatrix} -1 & 0 \\ 0 & 1 \end{pmatrix} - \begin{pmatrix} d_{11}E & d_{12}E \\ d_{21}E & d_{22}E \end{pmatrix}, \tag{9}$$

Under supposition that all relaxation processes can be neglected[16,17]:

$$i\hbar\frac{\partial \rho_{21}}{\partial t'} = -[\hbar\omega_a - (d_{22} - d_{11})E]\rho_{21} + d_{21}(\rho_{22} - \rho_{11})E, \tag{10.1}$$

$$i\hbar\frac{\partial}{\partial t'}(\rho_{22} - \rho_{11}) = 2d_{12}E\rho_{21} - 2d_{21}E\rho_{12}, \tag{10.2}$$

It is convenient to introduce the Bloch vector components

$$r_1 = \rho_{12} + \rho_{21}, \quad r_2 = -i(\rho_{12} - \rho_{21}), \quad r_3 = \rho_{22} - \rho_{11},$$

and to re-write equations (10) in the form, which we will call for shortness also Bloch equations:

$$\frac{\partial r_1}{\partial t'} = -[\omega_0 + (d_{11} - d_{22})E/\hbar]r_2, \tag{11.1}$$

$$\frac{\partial r_2}{\partial t'} = [\omega_0 + (d_{11} - d_{22})E/\hbar]r_1 + 2(d_{12}E/\hbar)r_3, \tag{11.2}$$

$$\frac{\partial r_3}{\partial t'} = -2(d_{12}E/\hbar)r_2, \tag{11.3}$$

Now we supplement them with the coupling equations (7)

$$E_{tr}(t) \equiv E(t') = \frac{2V_+}{V_- + V_+}E_{in}(t') - \frac{4\pi V_- V_+ n_A l}{c^2(V_- + V_+)}\frac{\partial}{\partial t'}\left\langle \frac{1}{2}(d_{22} - d_{11})r_3 + d_{12}r_1 \right\rangle, \tag{11.4}$$

where angle brackets mean a summation over all atoms of a thin film dividing by surface density of atoms $n_S$, $\omega_0$ is a central frequency of a statistical distribution of resonance frequencies $\omega_a$ of atoms.

The system of equations (11) can be re-written in terms of dimensionless variables:

$$\tau = \omega_0 t', \quad e = 2d_{12}E/\hbar\omega_0, \quad \mu = (d_{11} - d_{22})/2d_{12}, \quad \nu = (d_{11} + d_{22})/2d_{12}.$$

So that

$$r_{1,\tau} = -(1+\mu e)r_2, \quad r_{2,\tau} = (1+\mu e)r_1 + er_3, \quad r_{3,\tau} = -er_2, \tag{12.1}$$

$$e = Te_{in} + \kappa <r_2>, \tag{12.2}$$

where $e_{in} = 2d_{12}E_{in}/\hbar\omega_0$ is the normalized strength of a videopulse incident the interface,

$$T = 2V_+/(V_- + V_+)$$

is the transmission coefficient of the interface of two linear dielectric media,

$$\kappa = \frac{8\pi V_- V_+ d_{12}^2 n_A l}{\hbar c^2 (V_- + V_+)} = \frac{2v_- v_+}{(v_- + v_+)} \cdot \frac{l/c}{t_c}$$

is the coupling constant.

Here $t_c = \hbar/4\pi d_{12}^2 n_S$ is the cooperative time, $v_\pm$ the pulse velocities by both sides of the interface in terms of speed of light $c$. In derivation of (12.2) the consequence from Bloch equations (12.1): $\partial(r_1 - \mu r_3)/\partial\tau = -r_2$ was employed.

### 4. EFFECT OF LOCAL FIELD

The model considered above does not take into account the influence of the Lorentz local field. To make a simple generalization one, following[22,23], can do that by substituting the field acting upon atoms $E$ with the local field $E_L = E + \eta P$, where $P = (n_S/l)p$ is bulk polarization of the medium, and $\eta$ is the form factor responsible for specific features of the environment acting on a given atom. In bulk material this factor is $4\pi/3$.

In terms of normalized variables, the corresponded Bloch equations take the form:

$$\frac{dr_1}{d\tau} = -(1+\mu e_L)r_2, \quad \frac{dr_2}{d\tau} = (1+\mu e_L)r_1 + e_L r_3, \quad \frac{dr_3}{d\tau} = -e_L r_2, \tag{13.1}$$

$$e_L = Te_{in} + \kappa\langle r_2\rangle + \eta g <v+\mu> + \eta g <r_1 - \mu(r_3+1)>, \tag{13.2}$$

where $g = 2n_S d_{12}^2/l\hbar\omega_0$ is the ratio of the energy of dipole-dipole interaction to the energy of resonance transition[29]. It is accepted that a film is in parelectric phase, i.e. there is no macroscopic dipole moment in it. That means $<d_{11}>=0$. Then finally, the system of equation (13) for further calculation reads

$$\frac{dr_1}{d\tau} = -(1+\mu e_L)r_2, \quad \frac{dr_2}{d\tau} = (1+\mu e_L)r_1 + e_L r_3, \quad \frac{dr_3}{d\tau} = -e_L r_2, \tag{14.1}$$

$$e_L = Te_{in} + \kappa<r_2> + \gamma <r_1 - \mu(r_3+1)>, \tag{14.2}$$

where $\gamma = \eta g$ is introduced. At the output the field is calculated with except for local field effect

$$e_{tr} = e_L - \gamma <r_1 - \mu(r_3+1)> \tag{14.3}$$

Note that system (14.1) describes a precession of vector $r$ over the vector of effective field

$$\mathbf{\Omega} = \{-e_L,\ 0,\ 1+\mu e_L\},$$

determined in its turn by the local field (14.2) (fig.1):

$$\frac{\partial \mathbf{r}}{\partial \tau} = [\mathbf{\Omega} \times \mathbf{r}]. \qquad (15)$$

In the absence of the external pulse, the local field is not zero that means a permanent precession of Bloch vector and, as a sequence, a continuous radiation from the film. The dissipative processes, which are not accounted in this consideration, can drive Bloch vector to equilibrium state and the radiation ends.

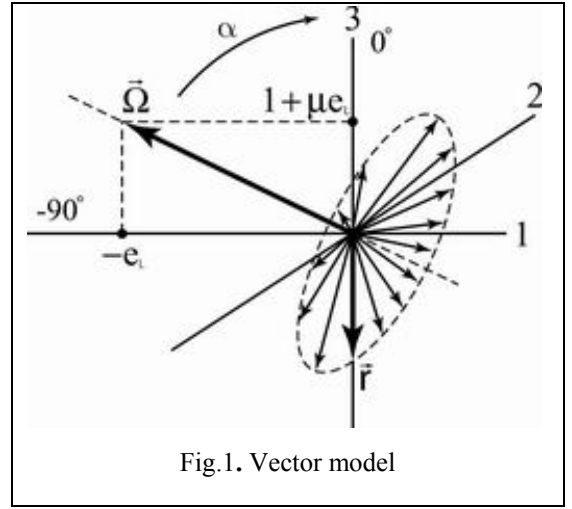

Fig.1. Vector model

## 5. NUMERICAL ESTIMATES

By choosing the magnitude of the dipole moment $d \approx 10^{-18}$ CGS, the density of resonance atoms $n_A \approx 10^{18}$ см$^{-3}$, and the duration of ultimately short pulse about $t_p \approx \omega_0^{-1} \approx 10^{-15}$s, one can get the estimate for the amplitude of the pulses $E_p \approx \hbar(2dt_p)^{-1} \approx 5 \times 10^5$ CGS at the peak intensity $I_p = cE_p^2/8\pi \approx 3 \times 10^{13}$ W/cm$^2$, and atomic field $\approx 10^6$ CGS.

The actual for two-level systems $t_c = \hbar/4\pi d^2 n_A \approx 8 \times 10^{-11}$s is the time of the formation of a dipole moment induced by the field of the traveling pulse. The normalized parameter $\tau_c = \omega_0 t_c \approx 8 \cdot 10^4$. During this time pulse passes the distance $L_{abs} = ct_c$ of about several cm with the spatial longitude of a pulse $ct_p \approx 3 \times 10^{-5}$ см. The value of $t_c^{-1}$ provides the estimation for Rabi oscillations in the characteristic feedback field of the medium $E_{char} \approx 2\pi n_A d = \hbar(2dt_c)^{-1} \approx 10$ CGS.

It is necessary to point out one more important relationship between the constants of the problem

$$\kappa = T v_{-} \frac{l/c}{t_c} = 4\pi^2 T v_{-} \frac{l}{\lambda_0} g, \text{ where } \lambda_0 = \frac{2\pi c}{\omega_0},\ g = \frac{2n_S d_{12}^2}{l\hbar \omega_0}.$$

As soon as $l \Box \lambda_0$ и $T v_{-} \approx T/n_{-} \approx 0.5$, then $\kappa \approx 4\pi^2 g \approx 20 g \approx 5\gamma$.

By choosing the thickness of the film $l \approx 10^{-6}$ m, for local field factor $\gamma$ and coupling constant $\kappa$ one gets

$$\gamma = \eta \frac{2n_S d_{12}^2}{l\hbar \omega_0} = \eta/2\pi \cdot \tau_c^{-1} \approx 8 \cdot 10^{-6}, \qquad \kappa = \frac{8\pi V_{-} V_{+} d_{12}^2 n_S}{\hbar c^2 (V_{-} + V_{+})} = \frac{2 v_{-} v_{+}}{(v_{-} + v_{+})} \cdot \frac{l/c}{t_c} = T v_{-} \frac{l}{L_{abs}} \approx 4 \cdot 10^{-5}.$$

As it is seen, the values of these key parameters are small, that does not provide a strong optimism to observe bright effects for the standard characteristics of resonance media. It is clear that the closer packing of atoms in film and larger dipole moments will increase the contribution of local field and enforce the observable effects. For instance, for larger dipole moments $d \approx 10^{-16}$ CGS, attributable for quantum dots, the cited parameters provide different orders of magnitude:

$E_p \approx 5 \cdot 10^3$ СГСЭ, $E_{char} \approx 6 \cdot 10^2$ СГСЭ, $t_c \approx 8 \cdot 10^{-13}$ с, $\tau_c \approx 8 \cdot 10^2$, $L_{abs} \approx 2 \cdot 10^{-2}$ см, $\gamma \approx 8 \cdot 10^{-4}$,

$\kappa = Tv_- \dfrac{l/c}{t_c} \approx 4 \cdot 10^{-3}$. The Stark parameter $\mu = (d_{11} - d_{22})/2d$ can be both positive and negative and has the magnitude $\mu \sim 0.2 \div 7.0$ for some sorts of semiconductor wells[12], $\mu \sim 0.4$, or, for example, for two low levels of vibration states of the ground state of molecule[13] $HeH^+$ or $\mu \sim 1.0$ by estimates from[11].

## 6. PHASE PLANE OF DYNAMICAL SYSTEM. ROBUSTNESS OF EQUILIBRIUM STATE

Let the spectral composition of the ensemble of permanent dipole atoms be homogeneous. The system (14.1) has an integral of motion standard for Bloch equation:

$$r_1^2 + r_2^2 + r_3^2 = 1. \qquad (16)$$

If to consider that the pulse $e_0(t) = Te_{in}(\tau)$ exhibits a very short, but strong impact upon a system, then, after this short "push" has passed, it can be supposed that the nonlinear system transfers to a state which differs from the equilibrium position at initial point $r_0 = (0,0,-1)$. Then vector $r = (r_1, r_2, r_3)$ moves in a manner prescribed by "free" ($e_0(t)=0$) equations:

$$\frac{dr_1}{d\tau} = -(1+\mu e_L)r_2, \quad \frac{dr_2}{d\tau} = (1+\mu e_L)r_1 + e_L r_3, \quad \frac{dr_3}{d\tau} = -e_L r_2, \quad e_L = \kappa r_2 + \gamma(r_1 - \mu(r_3+1)). \qquad (17)$$

Solutions (17) meet the trajectories on Bloch sphere fixed by expression (16). The Bloch sphere can be mapped on the plane of real variables (X,Y) by means of stereographic mapping (fig.2). Point A on the unit Bloch sphere is mapped on the (X,Y) plane in point A' with coordinates

$$x_{A'} = 2\cos\varphi \cdot ctg(\theta/2), \quad y_{A'} = 2\sin\varphi \cdot ctg(\theta/2).$$

If to choose mapping in the form $r_1 = \sin\theta\cos\varphi$, $r_2 = \sin\theta\sin\varphi$, $r_3 = \cos\theta$, then finally the mapping of Bloch sphere points on (X,Y) plane takes the view

$$x = \frac{2r_1}{1-r_3}, \quad y = \frac{2r_2}{1-r_3}, \qquad (18)$$

The north pole of the sphere is mapped (fig.2) in $\infty$, but the south pole converts in (0,0). The equations without external stimulations determine the rest points of (17). They look like following

$$(1+\mu e_L)r_2 = 0, \quad (1+\mu e_L)r_1 + e_L r_3 = 0, \quad e_L r_2 = 0,$$

$$e_L = \kappa r_2 + \gamma(r_1 - \mu(1+r_3)]) \qquad (19)$$

Equations (19) present a condition of co-linearity of vectors $\Omega$ and $r$ (15) in a stationary state of the system.

The state of rest $r_0 = (0,0,-1)$ for Bloch vector is the solution of (19). Under small initial deviations of Bloch vector from the south pole of the sphere (fig. 2) the state $r_0$ will be robust in the case of $\gamma < 1$.

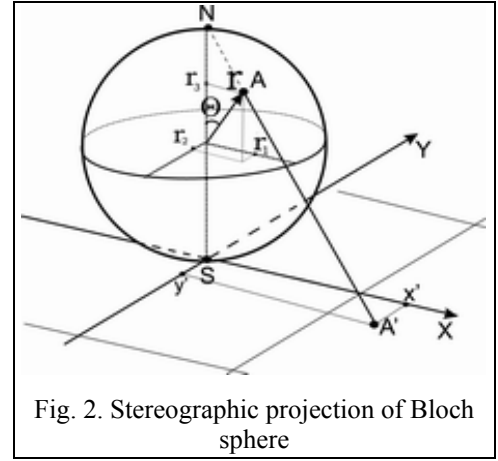

Fig. 2. Stereographic projection of Bloch sphere

If to linearize free equations (17) in the vicinity of $r_0 = (0,0,-1)$ by setting $r_1 = \delta r_1$, $r_2 = \delta r_2$, $r_3 = -1 + \delta r_3$, so that $e = \gamma \delta r_1 + \kappa \delta r_2 - \mu \gamma \cdot \delta r_3$, then for small values $\delta r_1, \delta r_2, \delta r_3$ one can get the linear system of equations

$$\frac{\partial \delta r_1}{\partial \tau} = -\delta r_2, \frac{\partial \delta r_2}{\partial \tau} = (1-\gamma)\delta r_1 - \kappa \delta r_2 - \mu \gamma \delta r_3, \frac{\partial \delta r_3}{\partial \tau} = 0. \qquad (20)$$

The equation follows from (20) is

$$\frac{\partial^2 \delta r_1}{\partial \tau^2} = -(1-\gamma)\delta r_1 - \kappa \frac{\partial \delta r_1}{\partial \tau} - \mu \gamma \delta r_3 (0). \qquad (21)$$

For $\kappa^2/4 < (1-\gamma)$ one obtains the oscillatory solution

$$\delta r_1(\tau) = -4\frac{\mu\gamma\delta r_3(0)}{(\omega^2+\kappa^2)} + \frac{4\exp(-\kappa\tau/2)}{(\omega^2+\kappa^2)}\left[\cos(\omega\tau/2)\left(\delta r_1(0)(\omega^2+\kappa^2)/4 + \mu\gamma\delta r_3(0)\right)\right]$$
$$-\frac{4\exp(-\kappa\tau/2)}{(\omega^2+\kappa^2)}\left[\sin(\omega\tau/2)\left(\frac{\omega^2+\kappa^2}{4\omega}(\kappa\delta r_1(0) - 2\delta r_2(0)) + \frac{\kappa\mu\gamma\delta r_3(0)}{\omega}\right)\right], \qquad (22)$$

It is seen from (22), that constant $\kappa$, responsible for the interaction of electromagnetic field with the medium of the layer, plays the role of the damping factor, but the quantity $4(1-\gamma)$ defines the square of eigenfrequency (without damping) of pendulum oscillation. The frequency of free oscillations is $\omega = \sqrt{4(1-\gamma) - \kappa^2}$. The solution of (17) for weak initial deviations of Bloch vector from the rest point is depicted in fig. 3. Under condition $0 \leq (1-\gamma) < \kappa^2/4$ one can obtain the solution of (21), where, in compare with (22), trigonometric functions are substituted for corresponding hyperbolic ones, and frequency $\omega$ for parameter

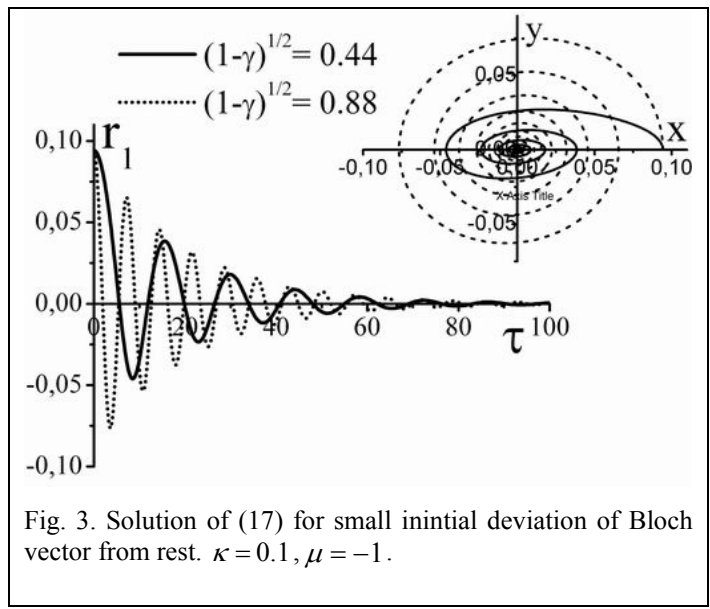

Fig. 3. Solution of (17) for small inintial deviation of Bloch vector from rest. $\kappa = 0.1, \mu = -1$.

$$\Delta = \sqrt{\kappa^2 - 4(1-\gamma)}.$$

The rest point remains stable, as the exponential factors will decrease faster than the hyperbolic functions with the argument $\tau\Delta/2$ grow.

What was said above is demonstrated in fig. 3(inset) where the trajectories in XY plane corresponding to the solution of (14) for small initial deviations from rest point are depicted. In fig.4a the trajectories conform to $\gamma = 0$ and 0.5 focuses in (0, 0). The corresponding solution is shown in fig. 5a. For $\gamma \approx 1$ the point of origin serves not as the focus, but the knot. The solution matches aperiodic damping. For $\gamma > 1$ the state of rest $r_0$ in the frames of linear analysis looses stability. The corresponding solutions of linearized problem grow with $\tau$. Generally speaking the linear analysis is no longer applicable.

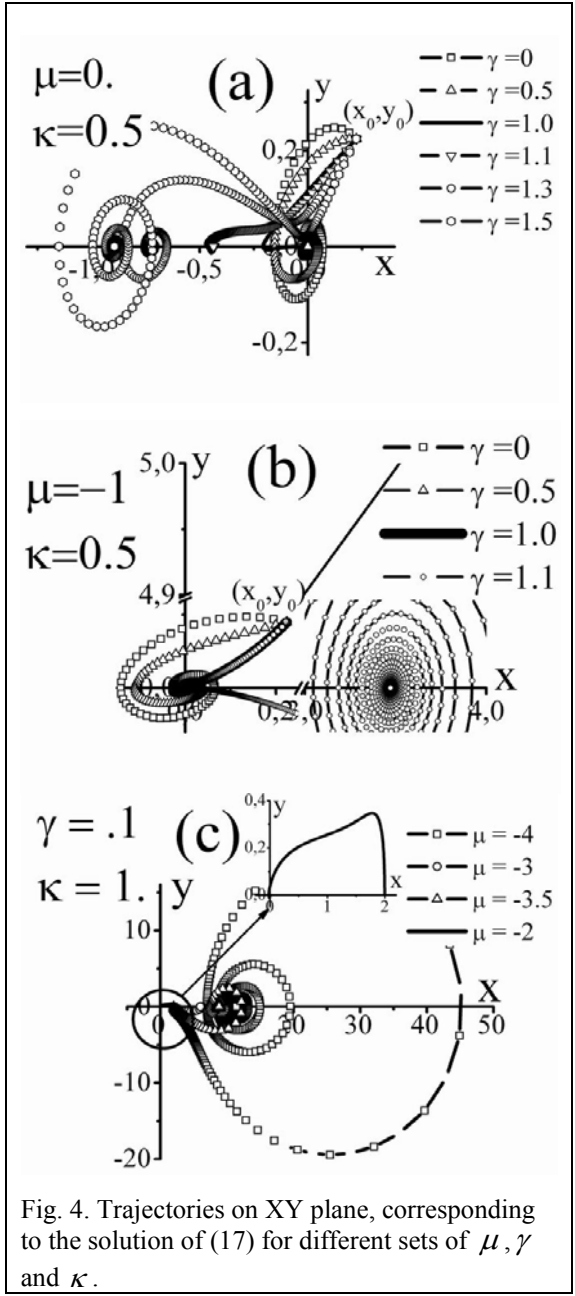

Fig. 4. Trajectories on XY plane, corresponding to the solution of (17) for different sets of $\mu, \gamma$ and $\kappa$.

## 7. NEW REST POINTS

Numerical modeling of the full, non-linearized system of equation (17) for not small initial deviations of matter variables demonstrates that trajectories associated with the solution of (17) find new rest points of the focus type including those when the inequality $\gamma > 1$ holds. The reason is the presence of a permanent dipole moment in atomic system causing the generation of local field which is zero only at the rest state $r_0$ of atoms. The variation of parameters $\gamma$ and $\mu$ affects the magnitude of local field and leads to the switch of the film transmittance even in absence of external field $e_0$, but in the presence of a noticeable deviation of Bloch vector from the rest state (fig. 4 a,b). The corresponding solutions of system (17) for film polarization and inversion tend to the different, nonzero values (fig. 5b). The non-zero negative values of Stark parameter $\mu$ effectively increase the local field $e_L$ (16). The fast oscillating damping solutions for this case are shown in figs. 4b and 5b.

Except $r_0 = (0,0,-1)$, equations (19) will be satisfied also at $e_L = 0$.

If $e_L \neq 0$, then the third equation of (19) demands $r_2 = 0$, then the first equation of the system is also satisfied. The rest equations look as following

$$(1 + \mu e_L)r_1 + e_L r_3 = 0, \quad e_L = \gamma[r_1 - \mu(1 + r_3)], \quad r_1^2 + r_3^2 = 1 \qquad (23)$$

Utilizing (18), equations (22) can be reduced to a single equation for variable $x$.

$$4x(4+x^2)(1-\mu^2\gamma) + 16x^2\mu\gamma + 4x\gamma(x^2-4)(1-\mu^2) - 2\mu\gamma x^2(x^2-4) = 0 \qquad (24)$$

Except the known root $x=0$, new rest points are determined by solutions $x(\mu)$ of equation (24), written in the form of polynomial:

$$F(x,\mu) = \mu\gamma^2 x^3 + 2x^2(2\gamma\mu^2 - 1 - \gamma) - 12\gamma\mu x + 8(\gamma - 1) = 0. \qquad (25)$$

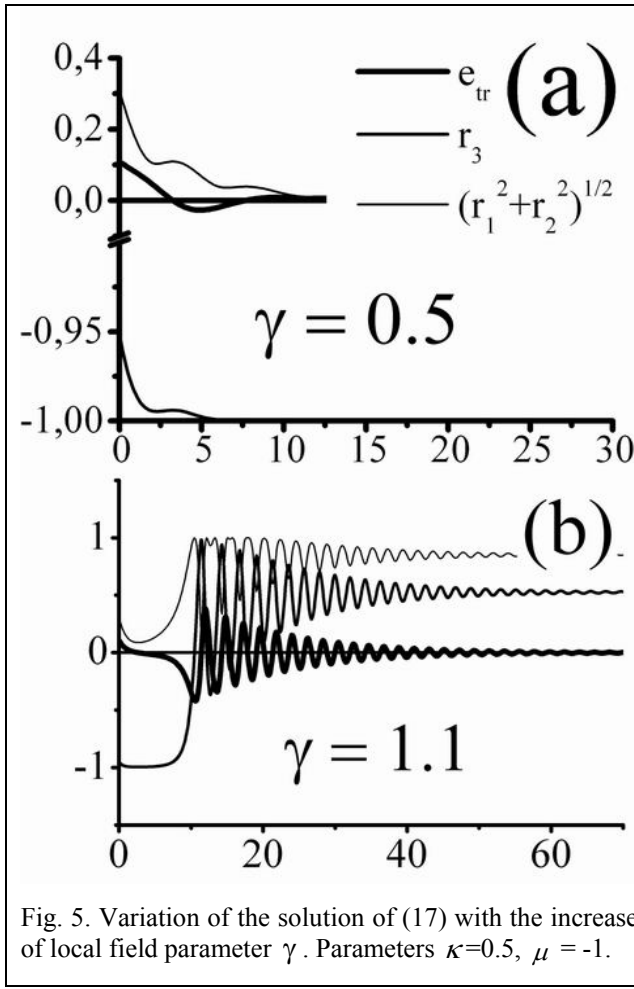

Fig. 5. Variation of the solution of (17) with the increase of local field parameter $\gamma$. Parameters $\kappa=0.5$, $\mu = -1$.

In order to obtain the graphics of function $x(\mu)$, one can plot the surface $F(x,\mu)$, and the zero-level line provides necessary curves. For some values of parameter $\gamma$ the plots of $x(\mu)$ are depicted in figs. 6. A central point on pictures corresponds to the solution $(x=0, y=0)$ also belonged to the set of roots of equation (23). The symmetry of the graphics presented on fig.6 follows from the symmetry of equation (24) for transformation $x \to -x$ and $\mu \to -\mu$. For $\gamma = 0$ equation (25) reduces to equation $x^2 + 4 = 0$, which has no real roots.

There is a gap between the branches of $x(\mu)$ for $\gamma < 1$ where there are no new rest points differ from $x=0$ (fig. 6 a,b) at least for moderate values of $\mu$. The specific features of these solutions are considered below. For $\gamma > 1$ the roots of equation (24) exist for any $\mu$. In particular, for $\mu = 0$ and $\gamma > 1$ equation (25) has the roots

$$x(0) = \pm 2\left(\frac{\gamma - 1}{\gamma + 1}\right)^{1/2}$$

(fig.6, c-f), who focuses the trajectories of system (17) associated with different initial conditions (fig. 6 d,e and fig. 7a,b). A stationary state (fig. 5b) corresponds to the focusing of the solution (fig. 4b) to $x$ coordinate, which is the root of (25) for $\mu = -1$ (fig. 6d)

# 8. THIN LAYER STATE SWITCHING

An existence of new stationary states in inversion and polarization of the layer of atoms possessing permanent dipole moment (fig. 4 (curves for $\gamma=1.1, 1.3, 1.5$), fig.5b) prompts that the analogy between the quantum object (14) (fig. 2) and a mechanical pendulum can be applicable even beyond the linear model (20) in solving a complete system (14) with the noticeable deviation of Bloch vector from an equilibrium state.

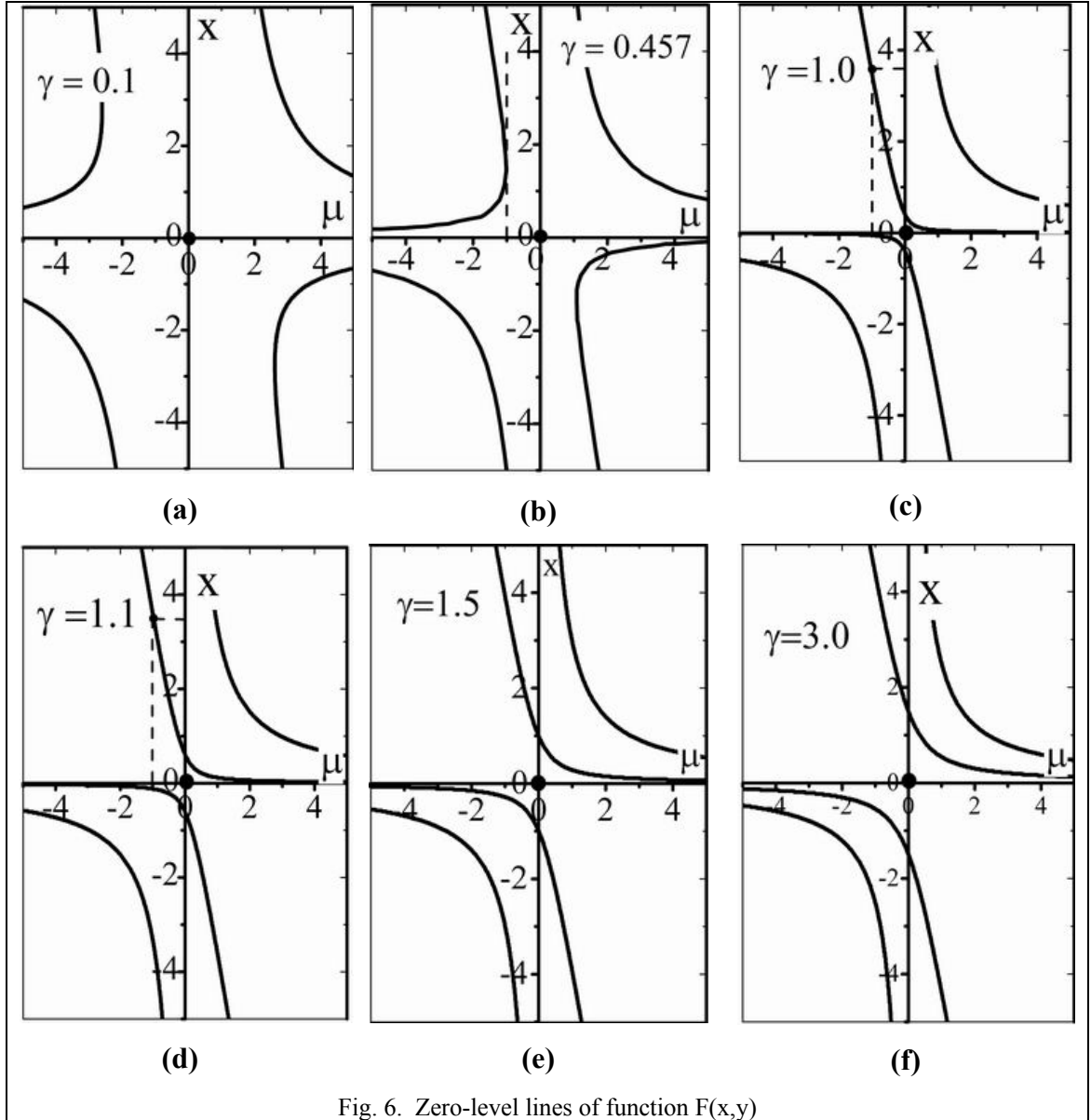

Fig. 6. Zero-level lines of function F(x,y)

So, the behavior of physical pendulum with the account of friction ($\beta$) and the moment of external force ($M$) conventionally looks:

$$\frac{d^2\psi}{d\tau^2} = -\omega^2 \sin\psi - 2\beta\frac{d\psi}{d\tau} + M(\tau),$$

where $\psi$ is the angle of deflection of a pendulum from vertical. With the increase of initial velocity $\psi'(0)$ up to the critical values (fig. 8a,b) the pendulum under the slightest alternation of initial ve-locity "switches" to one of stationary state: either it makes the revolution $\psi = 2\pi$, or it gets back to a stationary state $\psi = 0$.

The time slowdown (fig. 8a,b), characteristic for the solutions whose phase trajectories are close to separatrix, has the durations, which can be made very long by appropriate choice of initial velocity and by the increase of damping factor. Either short-term (fig. 8c), or permanent application of external force moment (fig. 8d) can switch over a pendulum to another stationary state. In the latter case, a constant moment of external force defines stationary position and this state does not equal to $2\pi$. In the problem under consideration the Bloch vector $r$ while precessing about the vector of effective field moves over the Bloch sphere like a physical pendulum, following the direction of $\Omega$.

The local field $e_L(\tau)$ (14.2) includes the incident external field $e_0$, the induced polarization $\kappa r_2$, and the proper internal local field $\gamma(r_1-\mu(r_3+1))$ is responsible for the effect of moment of external force. Under the low ini-tial velocities $\psi'(0)$ of pendulum, as well as for the small values of local field parameter $\gamma$, solution of (17) exhibits damping oscillations (fig. 3).

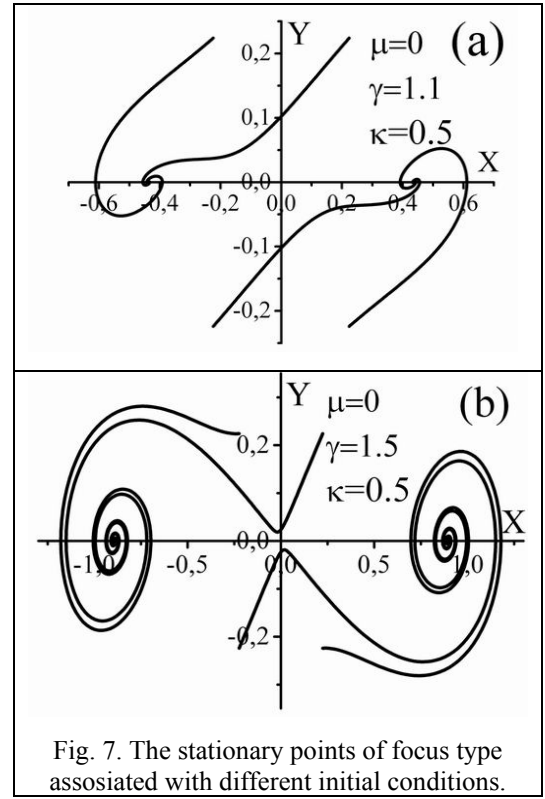

Fig. 7. The stationary points of focus type assosiated with different initial conditions.

At $\gamma > 1$ the rest state $r_0 = (0,0,-1)$, in the frames of linear analysis (21,22) looses its robustness, however the complete system of equations (14) possesses new, distinguished from $r_0$, stationary points. Under the minimal initial deviation from $r_0$ (external field is absent) the solutions changes barely noticeable over the prolonged interval like the motion in viscous medium.

On the XY-trajectory (fig. 9a (inset)) the slowdown is depicted by the closeness of dots. As soon as $r_3$ starts growing, the whole corrections to local field $\gamma(r_1-\mu(r_3+1))$ starts growing as well be-cause $\mu < 0$.

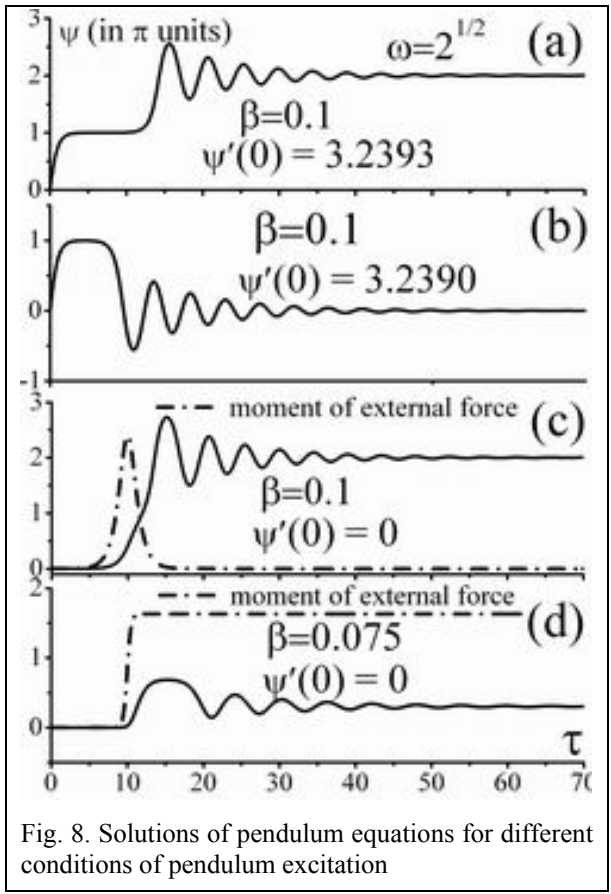

Fig. 8. Solutions of pendulum equations for different conditions of pendulum excitation

The quantum system shifts from unstable equilibrium and after the slowdown a quick switch over to a robust equilibrium state follows as the graphic on fig. 6c prescribes (fig.9a). The dynamics of quantum system reminds a transition of a pendulum to the new equilibrium state by application a moment of external permanent force (fig. 8d). The layer radiates a time-delayed oscillatory signal. The duration of delayed motion about the unstable equilibrium state shortens with the increase of both parameter $\gamma$ and the absolute value of Stark parameter $\mu$ because the local field grows. With the change of sign of Stark parameter $\mu$ and the decrease of $\gamma$ effect vanishes.

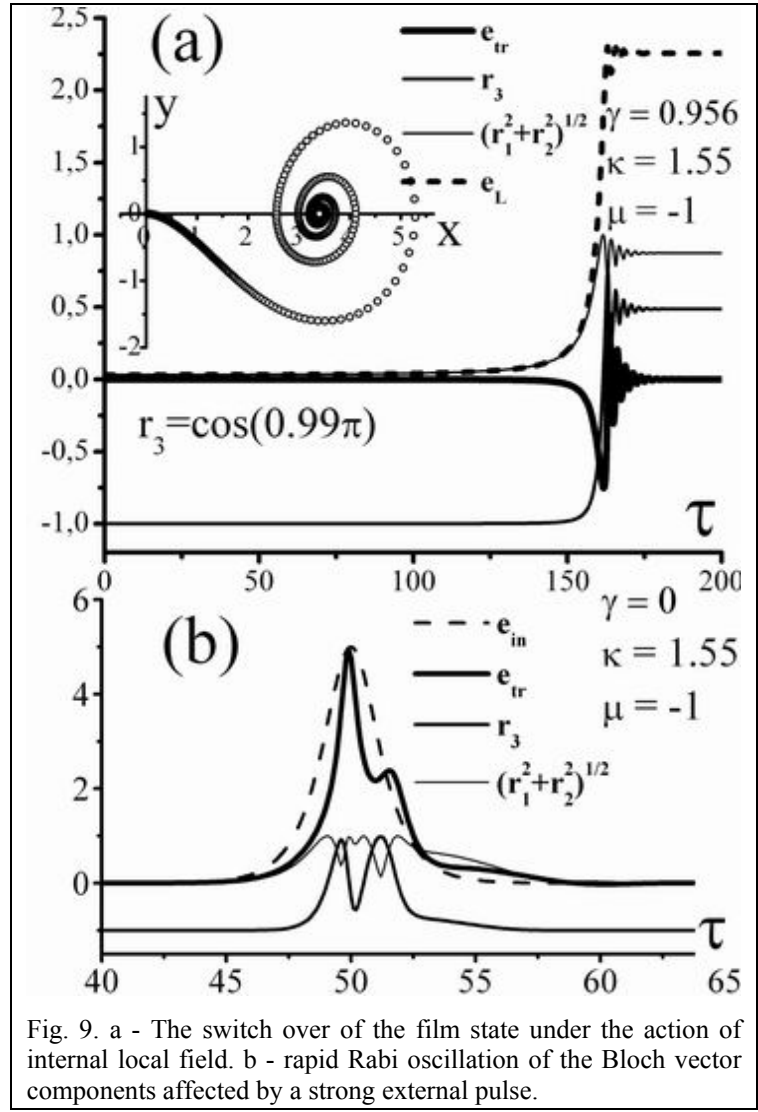

Fig. 9. a - The switch over of the film state under the action of internal local field. b - rapid Rabi oscillation of the Bloch vector components affected by a strong external pulse.

It is clear that the analogy of the quantum system under consideration to mechanical pendulum is not a detailed one. So, if a physical pendulum can be stimulated to make a turn by applying an appropriate moment of force of short duration (fig. 8c), one manage to transfer a quantum pendulum to a new stationary state only if $\gamma \neq 0$. Else the action of external field leads only to rapid Rabi oscillation of the Bloch vector components in pulse duration time (fig. 9b). At this time, a videopulse essentially acts as a trigger creating a seed inversion $r_3$ and polarization $r_1$ to generate internal local field.

Under non-zero values of the local field parameter $\gamma$ the probability of switch over from the main state $r_0$ to other stationary state depends in a threshold manner on the level of excitation (the values of $r_3$) the pulse has left the medium at. As soon as inversion oscillates in the field of videopulse, it can happen that to the end of the pulse atoms of the film are weakly excited and a switch does not occur even under the influence of a strong pulse or vice versa. However, in general, the rise of amplitude of incident pulse increases the probability of a switch in a layer transmittance. For instance, in fig. 10 parameter $\gamma$ is small and to observe a switch one should choose greater absolute values of Stark parameter $\mu$ (fig. 6a and fig.4c). The manifest change in film transmittance are well noticed in the difference between two phase diagrams of the processes (fig. 10a,c insets).

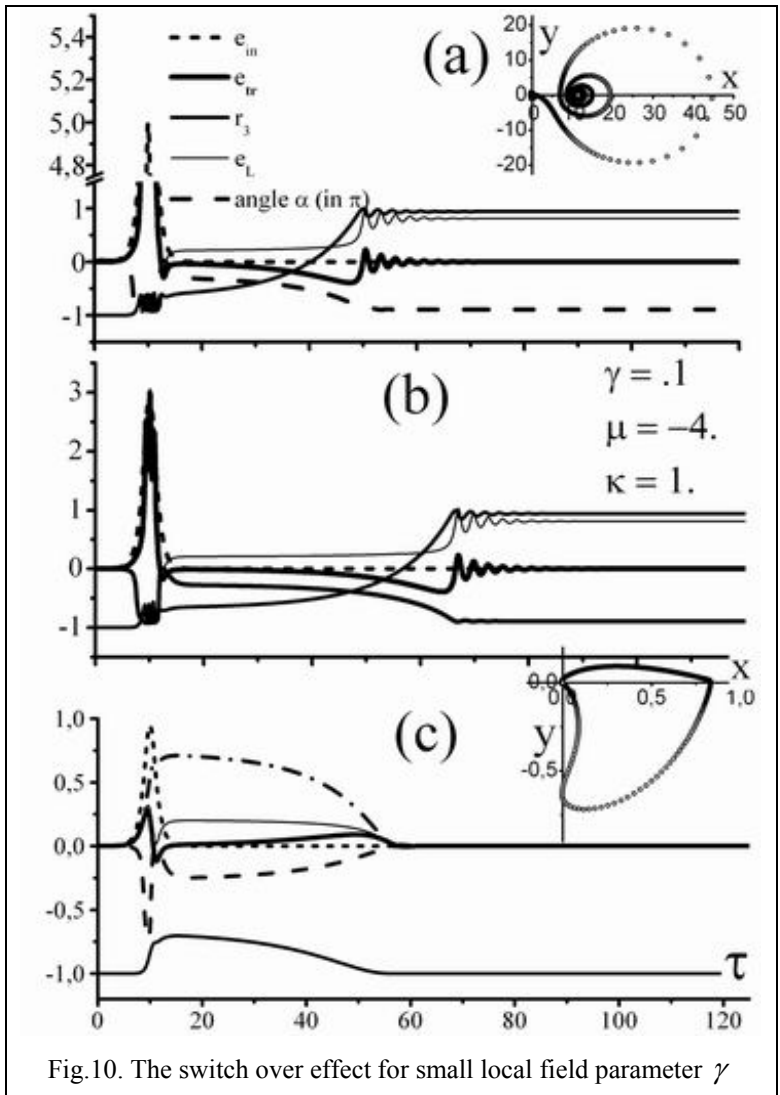

Fig.10. The switch over effect for small local field parameter $\gamma$

Naturally, the rotation angle of the effective field vector $\Omega$ (fig.1) goes down to $-\pi$ (fig. 10a), i.e. to the end of the process vector $r$ directs to the north pole of the Bloch sphere, whereas for the low levels of excitation the medium of a layer returns in the rest state with angle $\alpha$ goes to zero (fig. 10c).

In view of distinguished variance in the transition of atoms of the layer from one stationary state to another under the action of pulse field of different strength the dependence of transmittivity $\sqrt{W_{tr}/W_{in}}$, where $W$ is the pulse energy, against pulse amplitude experiences pulsations (fig.11) originated from the changes of film transmittance regime. With the increase of amplitudes of videopulse the transmittance tends Fresnel limit T=0.8 for current calculations. That means the film is getting more transparent for stronger pulses. This is the characteristic feature of other planar structure containing resonance particle[24].

As it was pointed out above under condition $\gamma < 1$ there are intervals of parameter $\mu$, where the only stationary point is – (x = 0,y = 0). It follows from the plots in fig. 6a,b, where, for example, the second stationary point appears as the intersection of the curve $x(\mu)$ with vertical line $\mu$ =-1 after the slightest increase of $\gamma$.

The trajectories of system (14) under such critical values of $\gamma$ pass nearby separatrix, that corresponds to the passage of pendulum close to the unstable equilibrium position. The speed of such motion is very low, thus the time to transfer to initial rest position, the delay time, (fig.12 lower panel) is very long (fig.13). The corresponding dependence of delay time vs $\gamma$ perfectly fits the reciprocal power law with the critical parameter $\gamma_c$ =0.4587.

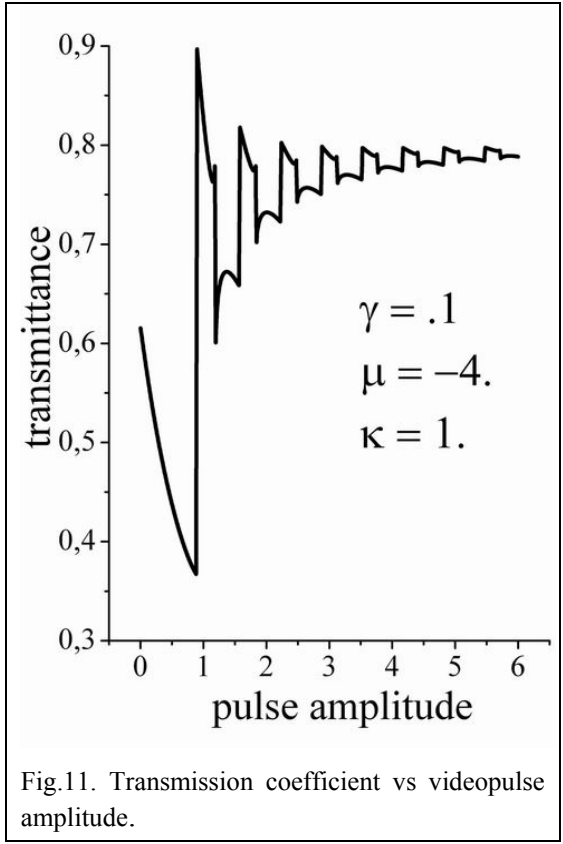

Fig.11. Transmission coefficient vs videopulse amplitude.

The position of stable equilibrium $r_0$ turns to be a knot (fig.12 upper panel), where solution gets over without oscillations. The decay of induced polarization while getting in stationary state causes an enlightening of a weak pulse of delayed radiation at the end of the Bloch vector motion (fig.12 lower panel). In fig. 12 the 3D picture on the middle demonstrates the temporal dynamics of the transmitted radiation under continuous alternation of local field parameter $\gamma$. The remarkable feature of this dependence is a fast, threshold appearance of the delayed signal when $\gamma$ approaches to critical value $\gamma_c$. The analogous threshold dependences in transmitted radiation take place under variation of $\mu$ and $\kappa$, as both parameters affect the local field strength.

## 7. CONCLUSION

In conclusion the propagation of videopulse through a thin layer of two-level spectrally homogeneous atoms possessing permanent dipole moments is numerically considered. The approximation of slow amplitudes and rotating waves are lifted. It turns out that the combination of three distinguish factors: the ultimate shortness of excited videopulse, the principal importance of local field effect in thin film physics, and the presence of permanent dipole moment in atomic subsystem gives rise to the effect of delayed pulsed electromagnetic radiation when the delay time is many times longer than all charachteristic temporal parameters in the problem under consideration.

An evident analogy with nonlinear mechanical pendulum permits to suppose that we meet the unique situation when local Lorentz field, which includes the effect of permanent dipoles, drives quantum system from one stationary state to another.

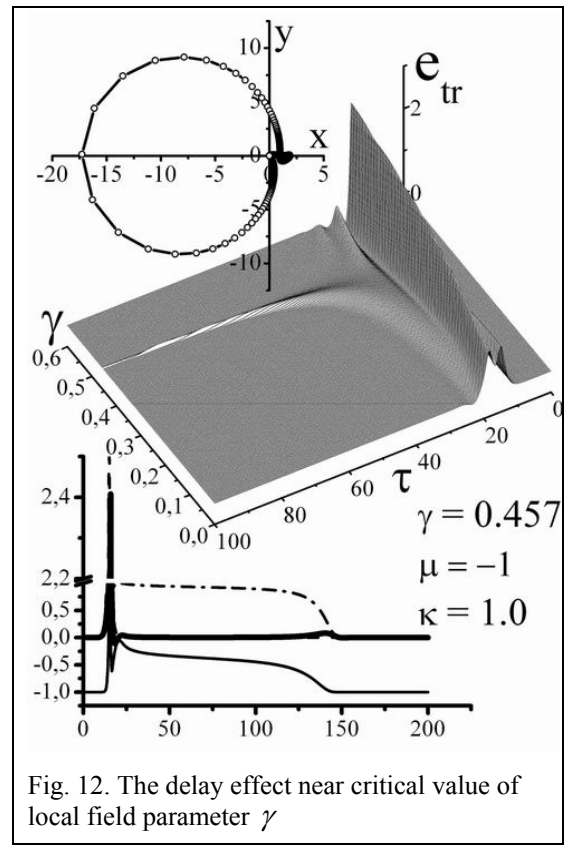

Fig. 12. The delay effect near critical value of local field parameter $\gamma$

A very long delay can be explained by a very slow motion of quantum system near unstable equilibrium. The effect has a threshold character over the strength of local field and depends upon the sign of Stark parameter $\mu$.

The integral curves of film energy transmittivity vs pulse field amplitude tends to Fresnel coefficient for given interface that agrees with a conventional result[24] when the stronger are the amplitudes of short pulses of electromagnetic radiation, the better a thin film of resonance particles transmits and worse reflects.

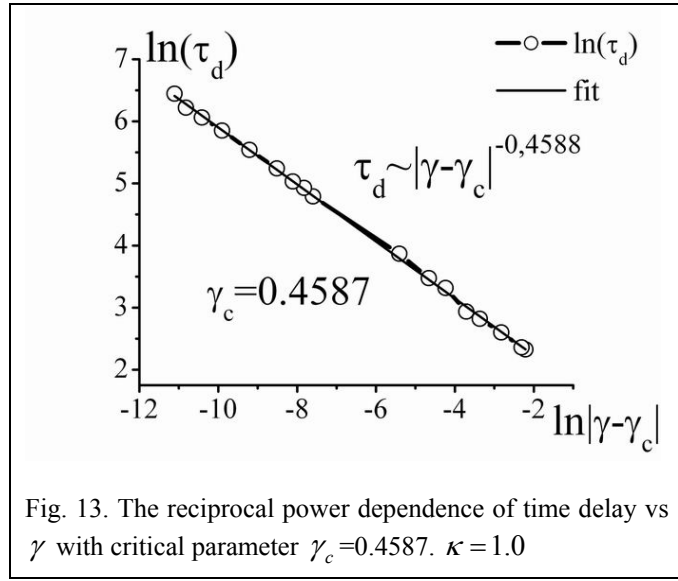

Fig. 13. The reciprocal power dependence of time delay vs $\gamma$ with critical parameter $\gamma_c = 0.4587$. $\kappa = 1.0$

## AKNOWLEDGMENTS

The authors gratefully acknowledge fruitful discussions with Askhat M. Basharov, and Jean-Guy Caputo. The research is supported by the RFBR grant 06-02-16406.